# Teaching Key Machine Learning Principles Using Anti-Learning Datasets

Chris Roadknight
School of Computing
*University of Nottingham*
Ningbo, China
chris.roadknight@nottingham.edu.cn

Prapa Rattadilok
School of Computing
*University of Nottingham*
Ningbo, China
prapa.rattadilok@nottingham.edu.cn

Uwe Aickelin
School of Computing
*University of Melbourne*
Melbourne, Australia
uwe.aickelin@unimelb.edu.au

*Abstract*—Much of the teaching of machine learning focuses on iterative hill-climbing approaches and the use of local knowledge to gain information leading to local or global maxima. In this paper we advocate the teaching of alternative methods of generalising to the best possible solution, including a method called anti-learning. By using simple teaching methods students can achieve a deeper understanding of the importance of validation on data excluded from the training process and that each problem requires its own methods to solve. We also exemplify the requirement to train a model using sufficient data by showing that different granularities of cross-validation can yield very different results.

## I. Introduction

Much of the teaching of machine learning revolves around using local knowledge, in N dimensional space, to iterate towards a global solution. This relies upon the premise that there will be a fitness 'hill' to climb on the way to the fitness 'peak' [4]. There are many examples of algorithms reaching local minima and methods to reset from or avoid these fitness sub-peaks [eg. 5], but overall there is an assumption that for supervised, semi-supervised and unsupervised learning methodologies that use local information will eventually reach the best possible solution and therefore provide the best possible model of the data.

There exists a range of problems where the use of local information to iteratively improve the accuracy of a model fail entirely. In fact, they fail so badly that the resulting model predicts categories in an unseen validation set at a rate much worse than guessing, this is called Anti-Learning [6]. Data that exhibits this kind of behavior comes from a range of sources both real and synthetic. Problems in the areas of cancer biomarker modelling, yeast genomics and chemical regulation are all real-world examples where there exists anti-learnable datasets. Synthetic data exhibiting this phenomena includes some traditional matrices and various versions of exclusive-OR data. This paper explores why this phenomena exists and why it is a crucial part of any machine learning curriculum, given it elegantly explains some key features such as the requirement for validation and the use of high granularity cross-validation.

## II. Anti-Learning

The meaningfulness of nearest neighbour style clustering in highly dimensional data has be discussed previously [2] and it has been argued that there are serious problems with using this approach alone [3], but this is exactly what happens in most undergraduate modules on machine learning.

Anti-learning is the situation when, no matter how hard you optimise your algorithms, the performance of your machine learnt model, on data not used for its training, is reproducibly worse than the probability of guessing the answer. So for a 2 class problem, if your model is consistently less than 50% accurate on unseen data, the underlying model needs to be anti-learnt. Most students, and even staff, are baffled by this scenario but its explanation ironically lies in one of the simplest, standard problems used to teach the need for multilayer perceptrons [1].

The exclusive-or problem is a trivial problem that is used by many teachers to explain why linear modelling solutions are insufficient to model non-linear problems [7]. It says that when inputs two inputs are both true or both false then the output should be false, while when the first and second input are different, the output should be true [17]. Figure 1 displays how the hyperplane needs to be separated (from [17])

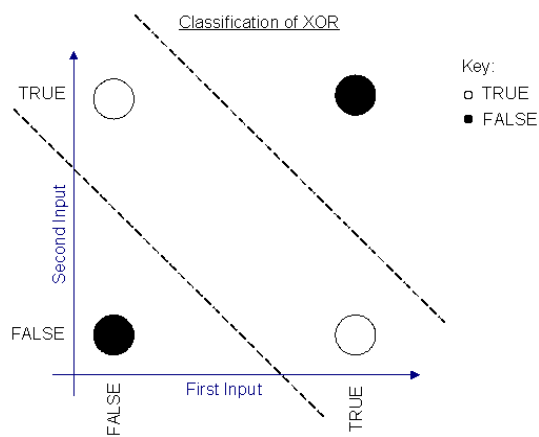

Fig. 1. Visual representation of exclusive-or

This problem requires a curved hyperplane to be generalised to accurately represent the data, and as such is a non-linear problem. The main stumbling block here though is that ALL of the dataset is required to build the model, so therefore there will be no data left to test the model. If you leave one value out of the training set, the model built will ALWAYS give the incorrect answer for these excluded value. For the Exclusive-Or problem where we know we have all available data, modelling using all the data is acceptable, if trivial but for most problems we only model the existing data in an effort to be able to approximate or predict outcomes for unrepresented or future scenarios, so therefore to test if we have generalized we need to assess performance on data not used to train the model.

### III. Data and tools

Three datasets are used in this work.

1. Hadamard matrix
2. An pyramidically aggregated Excusive-Or dataset
3. A randomly aggregated Exclusive-Or dataset

#### A. Hadamard matrix

A Hadamard matrix is a square matrix where each entry is either 1 or -1, and each pair of rows/columns has matching entries in exactly half of their columns and mismatched entries in the remaining columns/rows. The following is an example for hadamard matrices of size 2 and 4.

$H_2 =$  1,1
         1,-1

$H_4 =$  1, 1, 1, 1
         1,-1, 1,-1
         1, 1,-1,-1
         1,-1,-1, 1

Larger scale hadamard matrices can be displayed visually (fig 2).

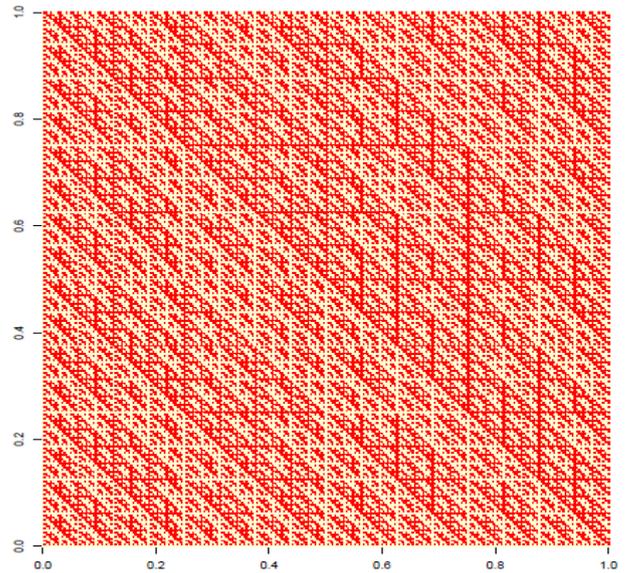

Fig. 2. Visual representation of a Hadamard Matrix

For the purposes of this work, we will use the proceeding n-1 values for each row as variables by which machine learning methods may attempt to learn dependent variable n (the final variable in the row).

For the purposes of teaching, this dataset is easily generated using a number of tools. For instance the function Hadamard(n) in the phangorn package [13] produces a hadamard square matrix of size n.

#### B. Pyramidically aggregated Exclusive-Or

A larger version of the standard exclusive-or logic can be generated by aggregating multiple XOR variable pairs. In this work we take 8 binary variables, a-h and calculate the XOR of the sequential 4 pairs, and then repeat this with the results. This is shown in equation 1.

$$y = xor(xor((xor(a,b),xor(c,d)), xor(xor(e,f),xor(g,h)))$$

(1)

This gives 256 training patterns that are exactly 50% True and 50% False.

Students can generate this dataset themselves using a variety of tools, including excel. The first few rows are shown below in table 1:

| a | b | c | d | e | f | g | h | out |
|---|---|---|---|---|---|---|---|---|
| 0 | 0 | 0 | 0 | 0 | 0 | 0 | 0 | FALSE |
| 0 | 0 | 0 | 0 | 0 | 0 | 0 | 1 | TRUE |
| 0 | 0 | 0 | 0 | 0 | 0 | 1 | 0 | TRUE |
| 0 | 0 | 0 | 0 | 0 | 0 | 1 | 1 | FALSE |
| 0 | 0 | 0 | 0 | 0 | 1 | 0 | 0 | TRUE |
| 0 | 0 | 0 | 0 | 0 | 1 | 0 | 1 | FALSE |
| 0 | 0 | 0 | 0 | 0 | 1 | 1 | 0 | FALSE |
| 0 | 0 | 0 | 0 | 0 | 1 | 1 | 1 | TRUE |

Table 1. First 8 rows of the aggregated 'pyramid' exclusive-or dataset

### C. Randomly aggregated Exclusive-Or

A more unbalanced version of aggregated XOR can be achieved by using a random variable structure. This has the benefit of allowing variables to be used multiple (or zero) times and is therefore a more heterogeneous structure. The random allocation we used is shown in equation 2.

$$y = xor(xor((xor(d,g),xor(a,d)), xor(xor(h,f),xor(d,b)))$$

(2)

This gives 256 training patterns that are exactly 50% True and 50% False. Again this dataset is trivial to produce using excel.

### D. Machine Learning

A multitude of off the shelf machine learning tools are available such as H2o [9], the caret [10] or e1071 [11] packages in R and WEKA [12]. A large range of algorithms have been used for anti-learning in the past [8] but here we just use Naïve Bayes, Support Vector Machine and a Multi-layer Perceptron.

## IV. RESULTS

We carried out performance tests on the three datasets, using three different machine learning algorithms and using 7 different levels of cross-validation granularity (4,8,16,32,64,128,256 folds). 256 fold cross-validation is equivalent to 'leave-one-out' (LOOCV) cross-validation. Figures 2, 3 and 4 show predictive accuracy on validation data. It is important to note that in all cases predictive accuracy on training data was always above 50% (usually above 75%).

The Naïve Bayes [16] approach (Figure 3) never predicts validation outcome at greater than 50% for any of the datasets at any of the cross-validation granularity. Moreover, at LOOCV it achieves 0% accuracy on all three datasets. Therefore by inverting the outcome Naïve Bayes can predict outcome on unseen data at 100% accuracy. This anti-learning result is better than any learning approach. The Multi-layer perceptron approach shown in figure 4 does achieve a high (>90%) accuracy on unseen data at low cross-validation granularity, but this falls swiftly to well below 50% once more cross-validation folds are introduced. This is a key result whereby the importance of cross-validation methodology is highlighted. The support vector machine [15] approach similarly tends towards an accuracy of well below 50% (figure 5) as the number of folds increases, though in this case the accuracy never goes above 50%, so is always in the anti-learning state.

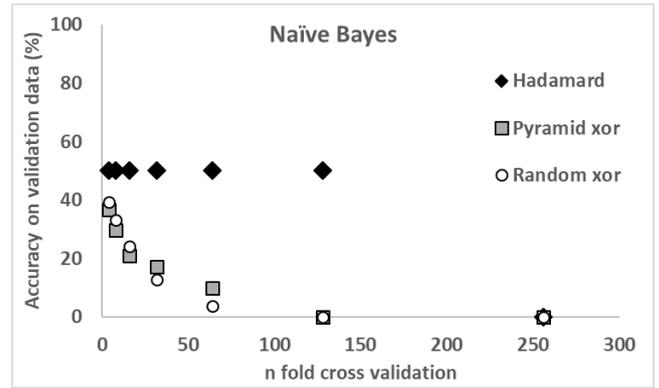

Fig. 3. Performace of a naïve Bayes approach on the three synthetic datasets for different granularities of cross-validation

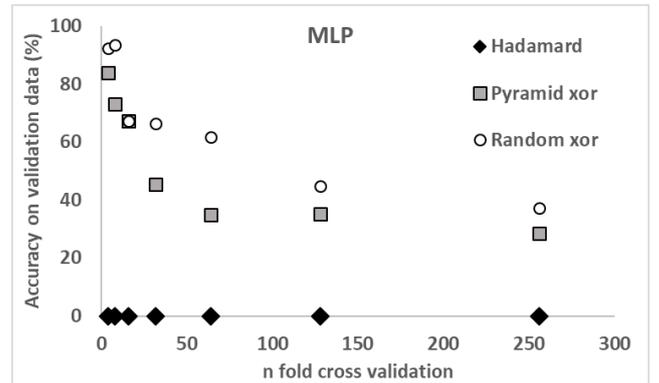

Fig. 4. Performace of a multi-layer perceptron approach on the three synthetic datasets for different granularities of cross-validation

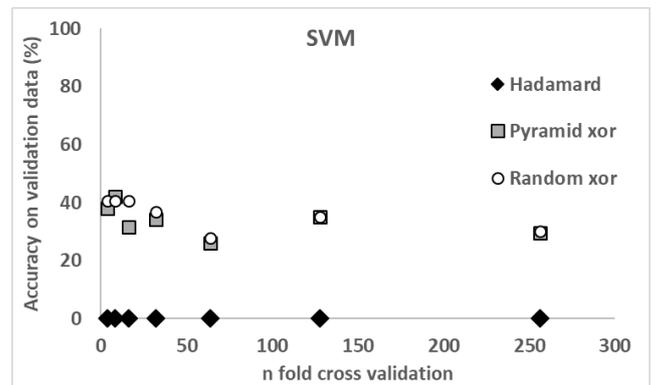

Fig. 5. Performace of a support vector machine approach on the three synthetic datasets for different granularities of cross-validation

## V. DISCUSSION

So why should anti-learning be included in undergraduate modules in machine learning. Most modules include supervised, unsupervised and semi-supervised learning paradigms in their curriculum. In this paper we suggest that anti-learning should be added to this. The reasons are as follows:

1. A small subset of real world and synthetic data are best modelled using an approach that exploits the anti-learning phenomena [8]. These real world problems are of high significance and complexity.

2. It elegantly teaches the importance of focusing on the performance of models on data that was excluded from the training process. Building a model that accurately reflects the data it is trained with is trivial but as is shown in this paper, largely meaningless if performance on unseen data (eg. future data) is a priority.

3. It highlights the importance that should be given to the validation process, in particular the degree of cross-validation. If it is not computationally impossible, 'leave one out cross-validation' is the safest approach, using the largest amount of data to train the models while still validating across the entire dataset. If the LOOCV results are very different to results using fewer folds, it must be suspected that an insufficient number of samples were used.

4. It teaches pragmatism. Some anti-learning datasets are highly dimensional with insufficient samples to represent the feature space completely but the data is still important and a timely model for subjects such as cancer prognosis is imperative. The best model turns out to be an anti-learnt model in some instances [14]. It may be possible to revisit the models at a later date, when more data is available.

5. It exemplifies the differences shown but different modelling techniques on the same dataset. In some instances choosing the most sophisticated, tunable modelling tool may not always be the best solution. Naïve Bayes is one of the oldest and simplest machine learning methods yet yields the most accurate anti-learning models for all three datasets.

Overall, introducing datasets that exhibit the anti-learning phenomena, whether it be because they are highly non-linear or they are highly dimensional, can only improve a student's understanding of appropriate methods of tackling all kinds of datasets.